# Algorithm for searching bridges $\overrightarrow{t^*}$ and $\overleftarrow{t^*}$ in the protection graph for Take-Grant protection model

## 1. Introduction

Take-Grant protection model is one of the earliest and most profoundly developed discretionary access control models [1]. One of the most significant achievements of Take-Grant model can be considered the ability to analyze system security in polynomial time. There are several papers that offer different variants for checking the security of computer systems, based on a Take-Grant model [2,3]. This paper is a continuation of [4,5] which describe ways of verifing the security of computer systems based on the terms set out in the classical Take-Grant model. This article describes a polynomial algorithm for searching bridges of type $\overrightarrow{t^*}$ and $\overleftarrow{t^*}$ in the protection graph. The proposed algorithm is based on a classical breadth-first search algorithm [6].

## 2. Initial conditions and notations

Based on the conditions formulated in the Take-Grant model, in order to investigate the possibility of access between the two subjects of an arbitrary protection graph, it is necessary that in the graph were known islands, bridges, initially and terminally spans of bridges. The islands are tg-connected subgraphs consisting only of subject vertices. Bridges, initially and terminally spans of bridges are paths of a given form, running through the object vertices. In this paper restrict the search of bridges $\overrightarrow{t^*}$ and $\overleftarrow{t^*}$ - paths in the graph, each arc of which contains a label $\overrightarrow{t}$ or $\overleftarrow{t}$, respectively.

Let there are already known island in the graph, a way to find them is described in [5]. It is necessary to find a bridge between two islands $I_1$ and $I_2$.

Introduce some notation. Arc between the vertices $e_i$ and $e_j$, containing the label $\vec{t}$, denote by $\vec{t}(e_i, e_j)$. Initial vertex of the bridge will be denoted by $s$, and the final – by $f$. The set of all object vertices of the original graph denote by $O$.

## 3. Bridge $\vec{t}^*$

Consider the search for a bridge of the type $\vec{t}^*$. At first we describe the algorithm informally. At the beginning the algorithm divides all set of object vertex of the original graph into two subsets $O_r$ and $O_i$. To the set $O_r$ falls those vertices to which there is a bridge of a given type, all other vertices falls in the set $O_i$. At the beginning of the algorithm in the set $O_r$ is only one vertex – $s$ (initial vertex). The algorithm looks at all the arcs of the graph which are associated with the vertices from $O_r$ and if it detect arc of type $\vec{t}$ which connects the top $e_r$ from $O_r$ with the top $e_i$ from $O_i$, then $e_i$ is removed from the $O_i$ and entered into $O_r$. After reviewing all the arcs there can be entered more than one vertex in the $O_r$, that is the cardinality of $O_r$ will increase, and the cardinality of $O_i$ will decrease by the same number. If the vertex $f$ falls in the set $O_r$, then it means that there is a path in the graph of a given form and the algorithm finishes its work. In other case, the procedure is repeated for the modified $O_r$ and $O_i$. However, it is possible that after reviewing all the arcs, to the $O_r$ will not be entered any vertices. This is possible when there are no arcs of the type $\vec{t}$ between vertices from $O_r$ and vertices from $O_i$. In order not to to miss this situation need to check the cardinality of the sets that we are dealing with. If the cardinality of the sets have not changed, it is necessary to finish the algorithm with a message that the specified type of bridge between these islands do not exist.

The formal algorithm is composed of three main steps. Before the beginning of the algorithm we divide the set $O$ into two subsets $O_r$ and $O_i$, i.e. $O = O_i \cup O_r$.

**Step 1.** Enter the vertex $s$ into $O_r$, enter all other vertices into $O_i$.

**Step 2.** Review all the arcs, which initial vertices are in $O_r$, if there is $\overrightarrow{t}(e_r, e_i)$, then enter $e_i$ into the $O_r$ and remove $e_i$ from the $O_i$. Here $e_r \in O_r$, $e_i \in O_i$. When all the arcs associated with the vertices from $O_r$ will be reviewed, go to **Step 3**.

**Step 3.** If after **Step 2** vertex $f$ is in the $O_r$, then the algorithm finishes – the bridge of the specified type exists. If after **Step 2** cardinality of sets $O_r$ and $O_i$ have not changed, the algorithm also finishes – the bridge of the specified type between these islands do not exist. Otherwise return to **Step 2**.

**Note.** It is possible to provide different implementations of the algorithm, depending on the task. For example, if there is only need to show the presence or absence of a bridge between the specified islands, then the above description would be enough - the algorithm reports the results of it's work. But if there is need to identify a bridge, than it is necessary to support the sets of passed arcs and vertices. For example, we can construct a graph of paths after each passage of the **Step 2**, or we can color passed vertices and arcs in some way.

Estimate the complexity of the algorithm. Let the original graph contains $N$ vertices. Since the graph directed, the maximum number of arcs in it may be equal to $N(N-1)$. It is possible to limit the number of repetitions of **Step 2** by the number of vertices, as in the case where after each step in the set $O_r$ will be entered only one vertex, then the bridge in a graph will be found in $N$ steps. In the general case there will be entered more than one vertex to the set $O_r$ for each execution of **Step 2**, that is, the bridge will be found for less than $N$ steps. If the bridge does not exist, than at some stage there will be no arc of the specified type and the algorithm does not entered in the set $O_r$ anything, ie, cardinality of the set $O_r$ will remain unchanged, then the algorithm will fail with appropriate message. Thus, at each **Step 2** there are required to review no more than $N$ vertices and with each vertex is connected by no more than *N(N-1)* arcs. That is, the complexity of the algorithm can be estimated as $O(N^3)$.

**Theorem.** The algorithm finds the bridge of type $\overrightarrow{t_*}$ correctly.

**Proof.** First we show that the algorithm finishes its work at all, and secondly, that the algorithm finds the right kind of bridge. We can show that the algorithm is finish it's work based on the fact that the set of object vertices of the original graph is finite. The algorithm divides the set $O$ into subsets $O_r$ and $O_i$, so that $O = O_i \cup O_r$. At each stage of **Step 2**, or any vertex is removed from the $O_i$ and entered in $O_r$ and the cardinality of both sets, respectively, change, or there are no movement of the vertices – the cardinality of the sets do not change and the algorithm fails.

Let $|O| = M$. There are three versions of events.

1. All the vertices of the $O_i$ will be transferred to $O_r$, then the algorithm is finished next step, as there is impossible to change the cardinality of the sets any more. Algorithm will make a total of *M+1* steps.

2. At step $k \leqslant M$ algorithm finds the bridge. The algorithm will finished it's work, and the number of steps it will make is equal *k*.

3. At step $k \leqslant M$ algorithm detects that the cardinality of $O_r$ and $O_i$ have not changed. The algorithm will finished it's work, and the number of steps it will make is equal *k*.

Thus, the algorithm is finished in any case, regardless of whether there are bridges in the graph or not.

The fact that the algorithm correctly finds the bridge, we can show by induction on the length of the bridge. Let the length of the bridge is *n*. We can not choose *n* = 1 as the induction basis, since this would mean that the initial and final vertices are connected by an arc $\overrightarrow{t}$, and therefore belong to the same island. Therefore, we choose n = 2 as basis and show that the algorithm finds this bridge.

The situation where there is a bridge of length 2 in the graph, is shown in

Figure 1. There is at least one arc of the type $\vec{t}$ associated with the vertex $s$, connecting vertices $s$ and $x$. In turn, there is at least one arc of the type $\vec{t}$ associated with the vertex $x$, connecting vertices $x$ and $f$.

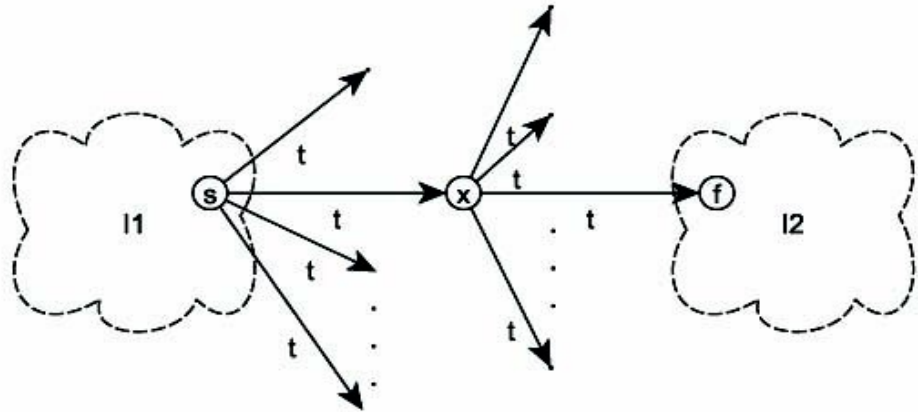

Figure 1 – Bridge $\vec{t^*}$ of length 2

The vertex $s$ entered in the set $O_r$ in **Step 1** of the algorithm. In **Step 2**, while checking all the arcs of type $\vec{t}$ associated with the vertices from the set $O_r$, the vertex $x$ will be detected. The vertex $x$ will be entered in the $O_r$ and removed from $O_i$, but the bridge has not been detected yet. The bridge will be detected only after repeated execution of **Step 2**, when the arcs of the type $\vec{t}$, associated with the vertices from $O_r$, will be reviewed again. This time will be found an arc connecting the $x$ and $f$.

As the induction hypothesis, we choose the statement that for the length of the bridge $n < l$, where $l > 2$, the algorithm finds the bridge correctly.

Inductive step: let the length of the bridge is equal $l$, the algorithm is executed $l-1$ stage, at this stage there were vertices $x_1, x_2, \ldots x_m$ entered in the $O_r$ (figure 2).

To each of the vertices $x_i$ bridge was found correctly by the induction hypothesis. Since the length of the bridge is equal $l$, this means that between at least one of the $x_i$ and $f$ there are arcs $\overrightarrow{t}(x_j, y)$ and $\overrightarrow{t}(y, f)$, that are the lsat arcs of the required bridge. We apply the algorithm for each of the $x_i$. The algorithm is able to find the bridge, consisting of two arcs, correctly as it was shown for the induction basis.

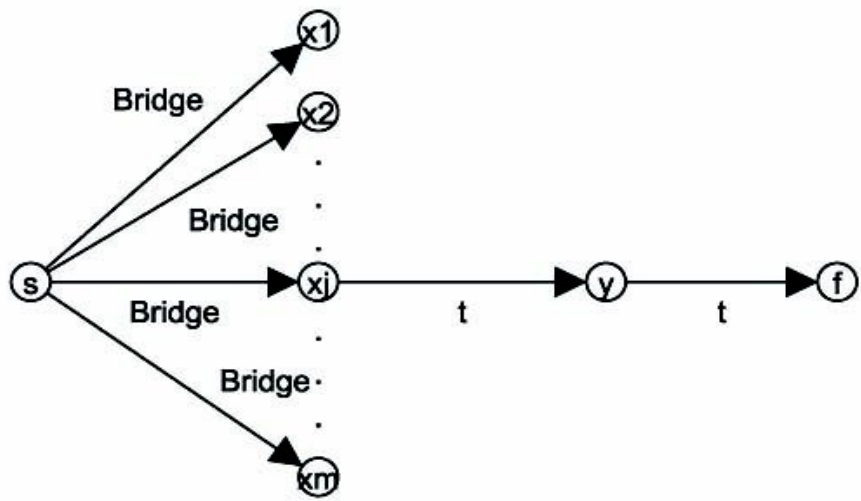

Figure 2 – Inductive step

**4**. Bridge $\overleftarrow{t}^*$

Obviously, if in the **Step 2** of the above described algorithm instead of the arcs of type $\overrightarrow{t}$ we will search arcs of type $\overleftarrow{t}$, than we can use this algorithm for searching bridge of type $\overleftarrow{t}^*$. In this case all of the above will be valid for a bridge of type $\overleftarrow{t}^*$, including the complexity of the algorithm will also be estimated as $O(N^3)$.

**5. Conclusions**

The search for bridges in the protection graph of Take-Grant protection model is needed to identify the channels of information leakage in a computer system. In order to find the channels of information leakage there are must also be ways to find bridges of type $\overset{\rightarrow}{t^*}\overset{\rightarrow}{g}\overset{\leftarrow}{t^*}$ and $\overset{\rightarrow}{t^*}\overset{\leftarrow}{g}\overset{\leftarrow}{t^*}$, as well as initially and terminally spans of bridges. Methods of searching these structures are not considered in this paper. However, the development of polynomial algorithms for searching these structures may form the basis for software safety analysis of computer systems. The algorithm described in this paper is just one step towards creating of such software.